\documentclass[11pt,a4paper]{article}

\usepackage{amsmath,amssymb,bm,mathtools}
\usepackage{balance}
\usepackage{graphicx}
\usepackage{tikz}
\usetikzlibrary{arrows.meta,positioning,calc,decorations.pathmorphing}
\usepackage[colorlinks=true,citecolor=blue,urlcolor=blue,linkcolor=blue]{hyperref}

\usepackage[titletoc,toc,title]{appendix}

\newcommand{\spll}{{/\kern-0.2em/}}
\newcommand\trick[1]{}
\newcommand{\be}{\begin{equation}} 
\newcommand{\ee}{\end{equation}}
\newcommand{\eq}[1]{(\ref{#1})}
\newcommand{\bit}{\begin{itemize}}  \newcommand{\eit}{\end{itemize}}
\newcommand{\ben}{\begin{enumerate}}  \newcommand{\een}{\end{enumerate}}

\def\bea{\begin{eqnarray}}
\def\eea{\end{eqnarray}}

\newcommand{\Order}{\mathcal O}

\def\Tr{{\rm Tr}}
\def\dd{\mathrm d}

\newcommand{\MP}{M_{\!P}}

\def\k{\kappa}

\def\O{\Omega}
 
\def\r{\rho}
  
\def\t{\tau}

\def\cD{{\cal D}}

\def\uno{\mbox{1 \kern-.59em {\rm l}}}

\def\bcomment#1{}

 \def\ii{{\rm i}}

\def\IR{\relax{\rm I\kern-.18em R}}

\long\def\symbolfootnote[#1]#2{\begingroup%
\def\thefootnote{\fnsymbol{footnote}}\footnote[#1]{#2}\endgroup}

\newcommand{\nthu}{{\it Department of Physics, National Tsing-Hua University,
    Hsinchu 30013, Taiwan}}

\newcommand{\ctc}{{\it
Center for Theory-Computation-Data Science Research, 
National Tsing-Hua University, Hsinchu 30013, Taiwan}}

\newcommand{\ncts}{{\it Physics Division,
    National Center for Theoretical Sciences, Taipei 10617, Taiwan}}

\begin{document}
\begin{center}
\vspace{20pt}
  
\thispagestyle{empty}
              {\Large \bf Quantum Horizon Tadpole and  Emergence of
                a de Sitter Interior}
                             
\vspace{25pt}

Chong-Sun Chu

\vspace{0.2cm}              

\vspace{5pt}\nthu\\
\vspace{5pt}\ctc\\
\vspace{5pt}\ncts

\vspace{1cm}

\begin{abstract}

In a recent paper \cite{Chu:2026dhx}, the quantum stability of the
fuzzy sphere was established at large but finite $N$, where a tadpole
was also identified for the scaling fluctuation mode of the matrix
geometry.  Here we show that this tadpole generates a positive
area-conjugate microscopic response, which in isolation energetically
favors contraction of the fuzzy sphere.  However, when the horizon is
coupled to gravity, the tadpole requires the bulk geometry to adjust
according to the junction condition of general relativity.  Taking the
exterior geometry to be Schwarzschild, we show that the junction
condition requires a positive interior cosmological constant. Imposing
a static, spherically symmetric, regular vacuum interior then selects
pure de Sitter space.  The matching further determines a soldering
relation between the interior de Sitter time and the exterior
Schwarzschild time, with a relative lapse growing as $C = O(N)$.  We
also propose a matrix realization of the de Sitter interior in which
fuzzy-sphere representation sizes determine the angular geometry while
condensates of off-diagonal link fields determine the emergent radial
metric. This provides a concrete consistency test for spacetime
emergence from the matrix degrees of freedom. Potential cosmological
implications are discussed.

\end{abstract}

\end{center}

\newpage
\setcounter{footnote}{0}

\tableofcontents
\section{Introduction}

Various mysterious properties of black holes have emerged over time
from attempts to understand the quantum aspects of an otherwise
classical black hole. Classic examples include the Bekenstein--Hawking
entropy, when one demands a statistical interpretation of black-hole
thermodynamics; the Hawking temperature and Hawking radiation, when
quantum matter fields are considered in a black-hole background; the
membrane paradigm, when one seeks an effective description of the
classical response of the horizon; black-hole chaos, whose classical
precursors can already be seen in the instability of near-horizon
trajectories; and, last but not least, the spacetime singularity in
the black-hole interior, whose resolution is expected to require
quantum effects of spacetime. These seemingly disparate phenomena
strongly suggest that a fundamental understanding of the microscopic
structure of the horizon is central to a consistent
quantum description of black holes.

Recently, a microscopic quantum-mechanical model of the black-hole
horizon has been proposed, in which the horizon arises as a solution
of a large-$N$ matrix quantum mechanics of quantum spaces \cite{Chu:2024qil}.
The model contains
three $N\times N$ traceless Hermitian matrix coordinates $X_a$, $a=1,2,3$, 
and two-component $N \times N$ matrix fermions
$\psi$.  $X_a$ transforms in the adjoint, while
$\psi$ transforms in the fundamental representation of
$SU(N)$ and so
the second matrix index of $\psi$ is a flavor index rather than a second index
of an adjoint representation.
The action  is 
\be \label{model}
 S={} \int dt\,  \Tr \Bigl[\frac{\dot X_a^2}{2a_0^2M_{\rm P}}
 +\frac{M_{\rm P}}{N^2}\bigl([X_a,X_b]^2+4X_a^2\bigr)
 +\ii\psi^\dagger\dot\psi
 -\frac{a_2M_{\rm P}}{N^2}\psi^\dagger\sigma_aX_a\psi\Bigr],
\ee
where $M_{\rm P}$ is a mass scale (Planck) of the theory, $a_0,a_2$ are
parameters that are fixed by requiring the theory to reproduce
the expected properties of black holes \cite{Chu:2024qil,Chu:2024edh}.
The
bosonic sector of the quantum horizon is described by a fuzzy sphere,
while the fermionic sector consists of a sea of fundamental fermionic
states quenched into the Lowest Landau Level (LLL) of an intrinsic
Berry monopole associated with the fuzzy sphere \cite{Chu:2026vzj}.
These fermionic
partonic states play several important roles: they account for the
Bekenstein--Hawking entropy through microscopic state counting
\cite{Chu:2024qil,Chu:2024edh};
the same states also select, through a matching of zero modes, the
tunneling path governing the non-perturbative decay of the fuzzy sphere
and thereby provide a microscopic description of Hawking radiation
\cite{Chu:2026wyh};
furthermore, they furnish the microscopic degrees of freedom
underlying a dynamical membrane paradigm for the quantum horizon
\cite{Chu:2026vzj,Chu:2026qom}.
Given the multiple roles played by the fermionic degrees of
freedom, it is natural to ask whether the bosonic quanta of the fuzzy
sphere likewise encode important aspects of the microscopic physics of
the horizon, and, if so, what these roles might be.

Work on this front has recently begun with a systematic
analysis of the bosonic fluctuations about the fuzzy sphere.
It was found that an interesting interplay between classical
and quantum effects takes place in the mesoscopic angular-momentum regime
$L \sim \sqrt{N}$. Positive quantum-corrected
curvature was established for all fluctuation modes,
while the special $(L,K)=(1,0)$ scalar sector develops a tadpole
\cite{Chu:2026dhx}.

The presence of a tadpole means that the chosen background is not an
exact stationary point of the quantum effective action and can drive
the theory toward a new quantum-corrected background, often with
interesting physical consequences. For example, in
non-supersymmetric string theory, a dilaton tadpole can similarly
induce a displacement of the string background through the
Fischler--Susskind mechanism \cite{Fischler:1986ci,Fischler:1986tb},
leading to a different spacetime
geometry.

In the present case, an important question is what the (1,0) scalar
tadpole does to the black-hole horizon. In this regard, we note that
the fuzzy-sphere horizon is never an isolated system. In the full
theory, it is embedded in an external environment described by
additional matrix blocks and interacts with the bulk degrees of
freedom through off-diagonal link blocks.  Such an enlargement of the
system already arose in our study of the electromagnetic membrane
paradigm \cite{Chu:2026vzj,Chu:2026qom},
where the off-diagonal links couple the quantum
horizon to the exterior electromagnetic field.
We should therefore consider the scalar tadpole in the complete
gravitating system consisting of the horizon and the bulk.

In the following, we show that the  (1,0) tadpole 
generates a positive area-conjugate microscopic response,
which in isolation
energetically favors contraction of the fuzzy sphere.
When gravity is included,  the gravitational background of
the coupled system should be fixed by the junction condition
\cite{Israel:1966rt,Barrabes:1991ng,Poisson:2002nv,Beltracchi:2021zkt}.
Taking the exterior background to be a Schwarzschild metric and utilizing
the discovered
microscopic area response,
we show that a  de Sitter metric is obtained for the interior
of the black hole. The matching also determines a soldering relation between the interior
de Sitter time and the exterior Schwarzschild time, with a relative lapse
that grows as $C=O(N)$.
This is strikingly different from the usual
vacuum picture of a Schwarzschild black hole. 
We further propose a
matrix realization of this interior geometry in which the representation
sizes of fuzzy-sphere blocks determine the angular radii, while
off-diagonal link condensates determine the emergent radial metric.
This leads to a concrete consistency test for the emergence of de Sitter
geometry directly from the matrix degrees of freedom.
Implications and further
comments will be made in the discussion section.

\section{Horizon Tadpole and Surface Area Response}
\label{tadpole}

We recall that the normalized (1,0) perturbation is described by
the vector
harmonics
\be  \label{B100}
 B^{100}_a
 =
 \frac{J_a}{\sqrt{D_N}},
 \qquad
 D_N:=
 \frac{N(N^2-1)}{4}.
\ee
The displayed configuration along this mode is therefore
\be
X_a=J_a+qB^{100}_a = \r J_a,
\label{rdef}
\ee
with
\be
\r :=1+\frac{q}{\sqrt{D_N}}.
\label{r-q}
\ee
Under the perturbation, the radius of the fuzzy sphere becomes
\be \label{Rq}
R_q = \r R,
\ee
where $R = N l_P$ is the radius of the original fuzzy sphere. 
Therefore, $\rho$ describes a scaling factor for the fuzzy sphere.
It was shown in \cite{Chu:2026dhx} that at large $N$, the 
one-loop corrected bosonic potential is given by
\be
V_{\rm eff}(q)
=V_{\rm eff}(0)
+\frac{\O_0}{2}A_{10}(N)\,q
+\frac{2\MP}{N^2}\Lambda^{\rm eff}_{10}(N)\,q^2
+\Order(q^3),
\label{master}
\ee
where
\be
  A_{10}(N) = 
\frac{8}{3}N^{3/2} + \Order(N^{1/2}),
\ee
and
\be
\Lambda^{\rm eff}_{10}
=4-\frac{8a_0}{N}+\Order(N^{-2}).
\label{maincoeffs}
\ee
Here $\O_0 := 2 a_0 M_P/N$ is a characteristic frequency unit for
the oscillation.
As a result, the quadratic curvature is positive for sufficiently large $N$.
The presence of the linear term means that $q=0$ is not an exact
stationary point of the quantum effective action.
Instead the tadpole induces, around the fuzzy sphere $q=0$, the
energy variation
\be
  dE_{\rm micro} =
 \frac{8a_0}{3}\sqrt{N}\,M_P dq.
\ee
The result is more informative when expressed in terms of the
geometrical area rather than the canonically normalized coordinate
$q$. Using \eq{Rq} for the scale deformation, we obtain
\be
 \left.\frac{dA}{dq}\right|_{q=0}
 =
 \frac{8\pi R^2}{\sqrt{D_N}},
\ee
and hence
\be \label{dEA}
dE_{\rm micro} = \tau_{\rm F}  dA
\ee
where
\be\label{t-FS}
\t_{\rm F} = \frac{a_0 M_P}{6 \pi l_P^2} \left(1+ \Order(\frac{1}{N}) \right)
>0.
\ee
We refer to $\tau_F$ as the microscopic area-tension coefficient, in
the sense that it is the coefficient conjugate to the area in
Eq.~\eq{dEA}.  Since $\tau_F>0$, increasing the area raises the
microscopic energy.  Thus, if this contribution were considered in
isolation, it would energetically favor a decrease of the fuzzy-sphere
radius.  Note that this contractive tendency of the
microscopic tadpole concerns the sign of the
microscopic energy variation and should not yet be interpreted as a
local mechanical surface stress.


Several features of this result are noteworthy.  First, although the
one-point function with respect to the normalized coordinate $q$
grows as $\sqrt N$, the area response $dA/dq$ has precisely the same
scaling.  Their ratio is therefore $O(N^0)$.  The resulting $\t_F$
is Planckian and, to leading order, independent of the macroscopic
radius $R=N\ell_P$.  This is a genuinely microscopic quantum response
of the horizon rather than a quantity suppressed by the size of the
black hole.
Second, the sign is fixed by the microscopic calculation.
Third, the tadpole is a first-variation effect and
should be distinguished from the quadratic fluctuation curvature,
which determines the local stability of the $(1,0)$ mode about a
configuration satisfying the full equations of motion.

The horizon is, however, not an isolated system.  In the full matrix
theory it is embedded in additional matrix degrees of freedom, with
the horizon block coupled to its environment through off-diagonal link
variables.  Once the fuzzy sphere is identified with a black-hole
horizon, its microscopic area response must therefore be considered
together with the gravitational response of the surrounding spacetime.
The relevant stationarity condition is not the vanishing of the
fuzzy-sphere tadpole by itself, but the vanishing of the total first
variation of the coupled system.  The nonzero tadpole should thus be
regarded as a physical source for the gravitational background rather
than something to be subtracted by hand.  Since Eq.~\eq{dEA} is
obtained entirely within the microscopic theory,
it provides a nontrivial consistency condition on any gravitational
embedding of the fuzzy sphere.

This is close in spirit to the Fischler--Susskind mechanism, in which a
nonvanishing quantum tadpole signals that the background about which
perturbation theory is performed must be shifted to one satisfying the
quantum-corrected equations of motion.  In the present case, if the
exterior Schwarzschild geometry is held fixed, the required
readjustment must be accommodated by the interior geometry.  The
microscopic area response may therefore provide a mechanism by which
quantum horizon dynamics selects an interior different from
the classical Schwarzschild one.

A useful analogy is provided by a domain wall in quantum field theory,
whose local properties are tied to the vacuum states on the two sides
that it separates.  Similarly, the fuzzy-sphere horizon may be viewed
as a {\it domain wall of quantum geometry} separating distinct exterior
and interior gravitational configurations.  One should therefore not
assume that a single Schwarzschild vacuum continues unchanged across a
horizon carrying the microscopic area response  \eq{dEA}, \eq{t-FS}.
The appropriate
framework for determining the geometries on the two sides and their
matching is provided by the junction conditions of general relativity,
to which we now turn.

\section{Null Junction Conditions and Horizon Energy}
\label{sec:nulljunction}

The fuzzy sphere is identified with a spatial section of the
black-hole horizon.  Its history in spacetime is therefore a null
hypersurface, rather than an ordinary timelike material shell.  The
standard Israel junction conditions \cite{Israel:1966rt}, which are
formulated for a non-null hypersurface, cannot consequently be applied
without modification.  We first recall the general form of the surface
stress tensor of a null shell \cite{Barrabes:1991ng,Poisson:2002nv}
and then specialize to the stationary, spherically symmetric
Killing-horizon junction \cite{Beltracchi:2021zkt} relevant to the
present problem.

\subsection{General null-shell stress tensor}

Let $\Sigma$ be a (2+1)-dimensional null hypersurface generated by a null tangent
vector $k^\mu$.  We choose coordinates $(\lambda,\theta^A)$ on
$\Sigma$, where $\lambda$ parametrizes the null generators and
$e_A^\mu$, $A=\theta,\phi$, span the spatial two-sections.  Since the
normal to a null hypersurface is itself tangent to the hypersurface,
the induced metric is degenerate.  It is therefore useful to introduce
an auxiliary transverse null vector $N^\mu$ satisfying
\be
 N_\mu k^\mu=-1,
 \qquad
 N_\mu e_A^\mu=0,
 \qquad
 N_\mu N^\mu=0 .
 \label{null-transverse-vector}
\ee
The discontinuity of the transverse curvature across $\Sigma$
determines the  stress tensor carried by the shell.

In the notation of Refs.~\cite{Barrabes:1991ng,Poisson:2002nv}, the
surface stress tensor of a general null shell can be written as
\be\label{general-null-stress}
 S^{\mu\nu}
 =
 \mu\,k^\mu k^\nu
 +j^A
 \left(
 k^\mu e_A^\nu+e_A^\mu k^\nu
 \right)
 +p\,\sigma^{AB}e_A^\mu e_B^\nu ,
\ee
where $\sigma_{AB}$ is the metric of the spatial two-section.
The three quantities appearing in Eq.~\eq{general-null-stress}
have distinct physical meanings.  The coefficient $\mu$ is the
surface energy density flowing along the null generators, $j^A$ is a
surface current, and $p$ is an isotropic tangential stress.  There is
no independent ordinary normal pressure, since the null normal is
itself tangent to $\Sigma$. For a spherically symmetric shell,
rotational invariance implies
\be
j^A=0,
\ee
although in general both $\mu$ and $p$ may be nonzero.

\subsection{Stationary spherical Killing horizons}
\label{SSKH}

The junction relevant to the present problem is more restrictive.
Consider a static, spherically symmetric geometry
\be \label{ds-fh}
ds^2=-f(r)dt^2+\frac{dr^2}{h(r)}+r^2d\Omega_2^2 .
\ee
A regular Killing horizon at $r=R$ has
\be
f(R)=h(R)=0 ,
\qquad
0<\lim_{r\to R}\frac{f}{h}<\infty .
\ee
For the Killing vector
\be
 \xi^\mu=(\partial_t)^\mu ,
\ee
the corresponding surface gravity
\be  \label{kappaBGM}
  \kappa
  =\frac12\sqrt{\frac{h}{f}}\,\frac{\dd f}{\dd r}\bigg|_{r=R}
\ee
is thus finite.

For a stationary spherical Killing horizon,
Ref.~\cite{Beltracchi:2021zkt} evaluates the singular part of the
Einstein tensor directly and finds
\be \label{mu-j-zero}
 \mu=0,
 \qquad
 j^A=0.
\ee
Thus the only nonvanishing surface components are tangential,
\be
\label{BGMstress}
S^A{}_B
=
\tau_s\,\delta^A{}_B,
\qquad A,B=\theta,\phi ,
\ee
where
\be
\label{tau-s}
\tau_S
=
\frac{[\kappa]}{8\pi G},
\qquad
[\kappa]:=\kappa_+-\kappa_- .
\ee
For $[\kappa]>0$, Ref.~\cite{Beltracchi:2021zkt} refers to
$\tau_s$ as the positive surface tension.  We emphasize, however,
that $\tau_s$ is the coefficient of the
null-horizon
stress tensor and should not at this stage be identified with the
microscopic coefficient $\tau_F$.  Their relation will be established
below through the gravitational energy variation.


We emphasize that the microscopic calculation of section \ref{tadpole}
does not
construct the null surface stress tensor \eq{general-null-stress}
directly.  In particular, we shall not identify $\tau_s$ with the
microscopic coefficient $\tau_F$ at this stage.  Instead, we will
demonstrate below how the
connection between the microscopic and gravitational descriptions can
be made through their response to an area variation.

\subsection{Komar energy and microscopic matching}
\label{sec:Komar}

For a stationary spacetime, the gravitational energy associated with
the Killing vector $\xi^\mu$ may be represented by the Komar charge
\be
\label{Komar}
E_{\rm K}[{\cal C}]
=
-\frac{1}{8\pi G}
\int_{\cal C}
dS_{\mu\nu}\,
\nabla^\mu\xi^\nu .
\ee
Using the Einstein equations, the difference of the Komar charges on
two closed surfaces can equivalently be written as
\be
\label{Komar-source}
E_{\rm K}[{\cal C}_2]-E_{\rm K}[{\cal C}_1]
=
2\int_{\Sigma}
\left(
T_{\mu\nu}-\frac12 Tg_{\mu\nu}
\right)
n^\mu\xi^\nu\,dV .
\ee
For a static diagonal geometry this takes the form
\be
\label{Komar-source-static}
E_{\rm K}[{\cal C}_2]-E_{\rm K}[{\cal C}_1]
=
\int_{\Sigma} d^3x\,\sqrt{-g}\,
\left(
-T^t{}_t+T^r{}_r
+T^\theta{}_\theta+T^\phi{}_\phi
\right).
\ee
Equation~\eq{Komar-source-static} is particularly useful for a
horizon junction.  The stress tensor contains a regular contribution
on either side of the horizon together with a
surface term.  The Komar energy therefore decomposes naturally into an
interior contribution and a horizon contribution,
\be
\label{Komar-decomposition}
M=E_V+E_S,
\ee
where $E_V$ denotes the Komar charge evaluated immediately on
the interior side of the horizon, while $E_S$ is the contribution
of the singular horizon stress.

For the stationary spherical null horizon of Section \ref{SSKH},
the singular
surface stress has only tangential components,
\be
S^\theta{}_\theta
=
S^\phi{}_\phi
=
\tau_S .
\ee
Its contribution to the Komar integral is therefore
\be
\label{Es-surface}
E_S
=
\int_H dA\,
\left(
S^\theta{}_\theta+S^\phi{}_\phi
\right)
=
2\tau_S A .
\ee
Thus the horizon contribution to the Komar energy follows directly
from the
tangential stress tensor.

The relation \eq{Es-surface} does not by itself imply that
$\tau_S$ is the coefficient of $dA$ in a general variation, since
$\tau_S$ may itself vary.  To make this distinction explicit, let
$\lambda_i$ denote the other independent data specifying the
junction.  We define the area-conjugate gravitational
coefficient $\tau_H$ by
\be
\label{dEs-general}
dE_S
=
\tau_H\,dA
+
\sum_i {\cal X}_i\,d\lambda_i .
\ee
The corresponding gravitational first law is then
\be
\label{firstlaw-general}
dM
=
dE_V
+
\tau_H\,dA
+
\sum_i {\cal X}_i\,d\lambda_i .
\ee

Now recall that
the microscopic calculation of Section 2 determines precisely an
area-conjugate variation for the microscopic fuzzy sphere energy,
\be\label{dEA2}
dE_{\rm micro}
=
\tau_F\,dA .
\ee
Now, since the tadpole corresponds to changing the radius of the fuzzy
sphere about a fixed background, with the remaining background data
held fixed, we may identify the microscopic energy variation
\eq{dEA2} with the area-variation term in the gravitational first
law \eq{firstlaw-general}.  This gives the matching condition
\be
\label{tension-match}
\tau_F=\tau_H .
\ee
We note that this
identification assumes that the microscopic and gravitational
energies are normalized with respect to the same time coordinate.
For the Schwarzschild junction considered below, this common
normalization follows from the previous matching of the matrix
Hamiltonian to the Schwarzschild mass.

Defined in this way, the gravitational coefficient $\tau_H$
as defined by \eq{firstlaw-general} is
logically distinct from the surface-stress coefficient $\tau_S$ of
\eq{Es-surface}. In general, the relation between
$\tau_H$ and $\tau_S$ is a property of
the particular gravitational junction and must be established
separately.
In a given gravitational solution, the two may coincide.
A familiar example is provided by the Schwarzschild--de Sitter
construction of Ref.~\cite{Mazur:2015kia}, where the Komar-energy
variation gives
\be 
dM=dE_V+\tau_S\,dA,
\ee
so that the area-conjugate coefficient coincides with the tangential
surface tension.  We shall show below that
\be
\tau_H=\tau_S
\ee
also holds for the more general spherical junction considered here.

\section{de Sitter Bulk Interior from Tadpole} 
\label{sec:interior}

We now apply the preceding results to the fuzzy-sphere black-hole
horizon.  We take the exterior geometry to be Schwarzschild and ask
which static, spherically symmetric interior geometry can be joined to
it across the horizon while reproducing the microscopic 
area response.

One feature of a Killing-horizon junction will be important below.
The stationary Killing vector
\be
\xi=\partial_t
\ee
is normalized by the asymptotically flat Schwarzschild region, and
this normalization will be kept fixed throughout the junction.  Thus
the freedom considered below should not be confused with a rescaling
of the Killing vector, or with the auxiliary null-basis transformation
$n^\mu\to\lambda n^\mu$, $N^\mu\to\lambda^{-1}N^\mu$.

Consider a junction with
interior and exterior geometry specified by
the  static, spherically symmetric metric \eq{ds-fh}. 
As emphasized in Ref.~\cite{Beltracchi:2021zkt}, although $f$ and
$h$ themselves match continuously to zero at the horizon, the finite
limiting value of $f/h$ need not be the same on its two sides.  This
allows a relative lapse normalization in the interior metric while
keeping the same asymptotically normalized Killing vector $\xi$.
Accordingly, we take
\be
ds_+^2
=
-f_+(r)dt^2+\frac{dr^2}{f_+(r)}+r^2d\Omega_2^2 ,
\ee
for the exterior, while for the interior
\be
ds_-^2
=
-C^2 f_-(r)dt^2+\frac{dr^2}{f_-(r)}+r^2d\Omega_2^2 .
\ee
Thus
\be
\left.\frac{f}{h}\right|_+=1,
\qquad
\left.\frac{f}{h}\right|_-=C^2 .
\ee
We define $C>0$ to be the positive relative lapse factor between the
interior and exterior metrics.  We choose the time orientation of the
interior Killing field to agree with that fixed by the asymptotically
normalized Schwarzschild Killing time, so that physical evolution on
the two sides proceeds in the same direction with respect to the
common coordinate $t$.  The parameter $C$ is therefore a
dimensionless metric datum of the junction, specifying the relative
interior lapse with respect to the fixed Schwarzschild time.  It is
not a rescaling of the normalized Killing vector or of the auxiliary
null basis
\footnote{This distinction is important in comparing with
Ref.~\cite{Mazur:2015kia}.  Appendix C of that reference rules out
rescalings of the auxiliary null vectors once the Killing field has
been normalized with respect to the asymptotic Schwarzschild time.
The lapse factor considered here is instead part of the interior
metric itself.  Indeed, the de Sitter--Schwarzschild solution of
Ref.~\cite{Mazur:2015kia} has a discontinuous ratio $f/h$, with
$(f/h)_-=1/4$ and $(f/h)_+=1$.  In our notation this corresponds to
$C=1/2$ and hence $|\kappa_-|=\kappa_+$.}.

We now show that, assuming the black-hole interior is a
spherically symmetric vacuum solution of the Einstein equation,
the interior region must have a positive cosmological constant.
If, in addition, the interior is required
to be free of singularity, pure de Sitter space is selected.
To proceed, we note that
the form of $f_-$ is constrained by physical requirements.
It is  natural to impose the requirement that, just like the exterior region, 
the interior is also a  vacuum solution of the Einstein equation.
The horizon therefore acts as a domain wall separating two
vacuum solutions
of general relativity. By spherical symmetry,
the most general static solution with a
cosmological constant is the Kottler metric,
\be\label{f-}
f_- (r) =1-\frac{2Gm}{r}-\frac{\Lambda r^2}{3}.
\ee
The common-horizon condition
\be
f_-(R)=0
\ee
gives
\be
\label{horizon-mLambda}
\frac{2Gm}{R}
=
1-\frac{\Lambda R^2}{3}.
\ee

Let us now evaluate both the interior and surface contributions to the
Komar energy directly.  Both sides of the junction are described using
the same Killing vector $\xi=\partial_t$, whose normalization is fixed
by the asymptotically normalized Schwarzschild time.  This gives the surface
gravities 
\be
\kappa_+
=
\frac{1}{2R},
\ee
and
\be
\kappa_-
=
\frac{C}{2}f_-'(R)
=
\frac{C}{2R}
\left(
1-\Lambda R^2
\right).
\ee
The tangential surface-stress coefficient is therefore
\be
\label{taus-general}
\tau_s
=
\frac{[\kappa]}{8\pi G}
=
\frac{1-C+C\Lambda R^2}{16\pi G R}.
\ee
The Komar charge evaluated immediately on the interior side of the
horizon $ E_V =
\frac{\kappa_- A}{4\pi G}$ gives
\be
E_V =
C\left(m -\frac{\Lambda R^3}{3G}\right) =
\frac{CR}{2G}
\left(
1-\Lambda R^2
\right).
\label{Ev-general}
\ee
The intermediate equation reveals
an interior Schwarzschild contribution proportional to $m$
and a cosmological constant contribution 
proportional to $\Lambda$. 

The horizon contribution $E_S$, on the other hand,
is obtained independently from the
tangential stress.
We have from \eq{Es-surface} and \eq{taus-general}
\be
E_S =
\frac{R}{2G}
\left(
1-C+C\Lambda R^2
\right).
\label{Es-general}
\ee
Adding the two contributions gives
\be \label{MEE}
E_V +E_S
=
\frac{R}{2G}
=
M ,
\ee
in agreement with the Schwarzschild Komar mass measured at infinity.

Note that the horizon condition \eq{horizon-mLambda}
leaves an overall length scale $R$ together
with dimensionless data characterizing the junction. 
It is convenient
to take these data to be $C$ and
\be
y:=\Lambda R^2 .
\ee
Indeed, when expressed in terms of the dimensionless coordinate
$x=r/R$, the horizon condition \eq{horizon-mLambda}
allows the interior metric function to be written as
\be
f_-(x)
=
1-\frac{1-y/3}{x}-\frac{y}{3}x^2 .
\label{fminus-y}
\ee
In this dimensionless form, the Kottler geometry depends only on
$y$, and not separately on $R$ and $\Lambda$.  Varying $R$ at fixed
$(C,y)$ therefore changes only the overall scale of the geometry,
while keeping its dimensionless background configuration fixed.
We
therefore identify  this with the gravitational variation corresponding
to the $(1,0)$ scaling mode of the fuzzy sphere.

With this parametrization, varying Eq.~\eq{Es-general}
and using \eq{MEE} gives
the gravitational first law
\be
\label{dEs-Kottler}
dM
=
dE_V
+\tau_S\,dA
+
\frac{R}{2G}
\left[
(y-1)dC+C\,dy
\right].
\ee
To compare this relation with the microscopic tadpole calculation, we
must also fix the normalization of the time coordinate, and hence of
the corresponding energy.  For the round fuzzy sphere, the energy of
the matrix model was matched to the Schwarzschild mass
\cite{Chu:2024qil}. Since this Hamiltonian generates
translations in the matrix time $t$, the matching fixes the
normalization of $t$ to be that of the asymptotically normalized
Schwarzschild Killing time.  Consequently,
the microscopic and
gravitational energy variations appearing here have the
same normalization.

As discussed above, the $(1,0)$ mode is a pure scaling deformation.
Its gravitational counterpart varies the overall scale $R$ while
holding the dimensionless background data $(C,y)$ fixed.  Therefore,
for the variation relevant to the tadpole,
\be
dC=dy=0,
\ee
and Eq.~\eq{dEs-Kottler} reduces to
\be
dM=dE_V+\tau_S\,dA ,
\ee
or equivalently,
\be
\left.dE_S\right|_{C,y}
=
\tau_S\,dA .
\ee
Comparing with the definition
\eq{dEs-general}, we obtain
\be
\label{tauH-taus}
\tau_H=\tau_S
=
\frac{[\kappa]}{8\pi G}.
\ee
This result is valid for the general Kottler interior and does not
require $\Lambda$ to be positive or $m$ to vanish.
Combining Eq.~\eq{tauH-taus} with the microscopic matching
condition \eq{tension-match}, we finally obtain
\be
\label{master-junction}
\tau_F
=
\tau_H
=
\tau_S
=
\frac{[\kappa]}{8\pi G}.
\ee
Thus the equality between the microscopic tadpole coefficient and the
tangential null-horizon stress follows from the microscopic area
response, the Komar decomposition, and the gravitational first law.
It is important to distinguish the equality of the coefficients in
Eq.~\eq{master-junction} from an equality of their local mechanical
interpretations.  The microscopic coefficient $\tau_F$ is defined by
$dE_{\rm micro}=\tau_F\,dA$,
so that $\tau_F>0$ means that the microscopic contribution, considered
in isolation, energetically favors a decrease of the area.  By
contrast, $\tau_S$ is a component of the null-horizon surface stress
tensor and has the corresponding gravitational mechanical
interpretation.  Their equality should therefore be understood as an
equality of the area-conjugate coefficients in the microscopic and
gravitational energy variations, rather than as an identification of
their local mechanical interpretations.

Using the explicit surface gravities,  Eq.~\eq{master-junction}
gives
\be
\label{Lambda-C}
\Lambda R^2
=
1+\frac{16\pi G R\tau_F-1}{C}.
\ee
Since $C>0$, $R=Nl_P$, and the microscopic area response $\tau_F$ is
positive and finite at large $N$, we have
\be
16\pi G R\tau_F=O(N)>1
\ee
for sufficiently large $N$.  It follows directly that
\be
\Lambda R^2>1,
\ee
and hence
\be
\Lambda>0.
\ee
Thus a positive cosmological constant is not assumed, but follows from
the positive microscopic  area response and the null-horizon
junction condition. Moreover, $\k_- <0$ and so
the horizon bounding the interior is necessarily of cosmological
type.

We next consider the different possibilities of $m$. The case $m<0$
can be rejected immediately since it gives a naked singularity.
For $m=0$, Eq.~\eq{horizon-mLambda} gives
\be
\Lambda R^2=3,
\ee
and hence
\be
f_-(r)=1-\frac{r^2}{R^2}.
\ee
The interior is therefore pure de Sitter and is regular and static
throughout $0\leq r<R$.
For $m>0$, Eq.~\eq{horizon-mLambda}, together with
$\Lambda R^2>1$, gives
\be
1<\Lambda R^2<3.
\ee
The zero at $r=R$ is therefore of cosmological type,
$\kappa_-<0$, and $f_-(r)>0$ immediately inside the horizon.
On the other hand,
\be
f_-(r)\longrightarrow-\infty
\qquad r\longrightarrow0
\ee
because of the Schwarzschild term.
Hence $f_-$ must vanish again at some $r=r_b<R$.  This second zero is
a black-hole horizon, and the static region lies between $r_b$ and
$R$.  The geometry therefore contains another black-hole interior
rather than terminating at a regular center.
Thus, if the interior is required to consist of a single static
vacuum region that is regular all the way to the center, the $m>0$
branch is excluded and $m=0$ is uniquely selected.  If this
single-region requirement is relaxed, one may instead ask whether the
same junction mechanism operates at the inner black-hole horizon.  If
so, the construction may be iterated, leading to a nested sequence of
static regions.  We note, however, that the soldering data need not be
the same from one junction to the next.  Determining these data
therefore requires a separate matching at each stage, which we do not
pursue here.
For concreteness, we shall consider here the simplest case of
a pure  de Sitter space interior.
In this case, the soldering condition reads
\be \label{CN=}
C  = \frac{4\pi N}{9b} + O(1) = O(N).
\ee
Here we have used $a_0 = \pi/3$ and  the relations for $l_P, M_P$
in terms of the Newton constant $G$,
\be
l_P = \sqrt{\frac{2G}{\pi}}, \qquad
M_P = \frac{1}{b}\sqrt{\frac{2}{\pi G}},
\ee
with $b :=a_2 -1>0$. 
 
As a result, the tadpole backreacts on the bulk gravitational
background in such a way that the usual singular black-hole interior
is replaced by a non-singular de Sitter space. de Sitter--black-hole
junctions have been studied extensively in the past as models for
resolving the
black-hole singularity. See for example
\cite{Frolov:1989pf,Dymnikova:1992ux,Hayward:2005gi,Visser:2003ge,
  Mazur:2004fk,Mazur:2015kia,Beltracchi:2018ait}.
  In such constructions,
however, the matter
localized at the junction and its stress tensor are typically
introduced phenomenologically, with suitable properties assumed in
order to support the desired geometry. The situation is different
here: the coefficient of the required horizon stress is
fixed by the same microscopic horizon system,
governed by the same matrix dynamics, that have
already been used to account for other quantum properties of the black
hole. No additional shell matter or phenomenological equation of state
needs to be introduced. Thus the de Sitter interior is not supported
by an independently postulated material shell, but emerges from the
backreaction of the microscopic quantum horizon degrees of freedom
themselves.

\section{Matrix Realization of the de Sitter Interior}
\label{sec:matrix-dS}

The junction analysis above determines the gravitational geometry on
the interior side of the fuzzy-sphere horizon.  It is therefore natural
to ask how this geometry may be represented directly in the matrix
description.  In particular, we would like to formulate a criterion
that distinguishes a genuine dynamical realization of de Sitter space
from a mere parametrization of a collection of fuzzy spheres.

A constant-time slice of the de Sitter static patch is
\be \label{dS-spatial-r}
 ds_{\rm spatial}^2
 =
 \frac{dr^2}{1-r^2/R^2}
 +
 r^2d\Omega_2^2 .
 \ee
With a change of variable,
\be \label{r-rho}
r
=
R\sin\left(\frac{\rho}{R}\right).
\ee
a constant-time slice of the de Sitter static patch can be written as
\be \label{dS-spatial-rho}
 ds_{\rm spatial}^2
 =
 d\rho^2
 +
 R^2\sin^2\left(\frac{\rho}{R}\right)d\Omega_2^2,
 \qquad
 0\leq\rho\leq\frac{\pi R}{2}.
\ee
Thus a spatial slice is a hemisphere of a three-sphere foliated by
two-spheres whose radius increases from zero at the center to $R$ at
the cosmological horizon. 
This suggests a natural matrix realization
in terms of a sequence of fuzzy two-spheres.  Consider a direct sum of
fuzzy spheres with ranks $N_s$
\be
 {\cal H}_{\rm dS}
 =
 \bigoplus_s V_{j_s},
 \qquad
 N_s=2j_s+1 ,
 \ee
 where each block
\be
 X_a^{(s)}
 \simeq
 J_a^{(N_s)} 
 \ee
 describes a fuzzy two-sphere with  physical radius
\be \label{rs-Ns}
 r_s
 =
 l_P\sqrt{N_s^2-1}
 \simeq
 N_s l_P ,
\ee
at large $N_s$.  Thus the representation sizes $\{N_s\}$ determine the
angular radii of the spherical leaves.  The outermost block has
$N_s=N$ and $r_s\simeq R=Nl_P$, and is identified with the
fuzzy-sphere horizon considered in the preceding sections.

The direct sum by itself, however, describes only a disconnected
collection of fuzzy spheres and does not yet define a
three-dimensional geometry.  The radial structure must be supplied by
the off-diagonal matrix elements connecting the blocks.  Schematically,
we consider
\be \label{matrix-dS-block}
 {\mathbb X}_a
 =
 \begin{pmatrix}
 X_a^{(1)} & W_a^{(1)} & 0 & \cdots \\
 W_a^{(1)\dagger} & X_a^{(2)} & W_a^{(2)} & \cdots \\
 0 & W_a^{(2)\dagger} & X_a^{(3)} & \cdots \\
 \vdots & \vdots & \vdots & \ddots
 \end{pmatrix}.
 \ee
 The rectangular matrices $W_a^{(s)}$ connect neighboring
representations and play the role of bifundamental link fields.  For
blocks of different ranks, their angular dependence is naturally
described by fuzzy monopole harmonics, with monopole
charge determined
by the difference between the two representations.
A first requirement for the emergence of a radial geometry is therefore
that the link condensates dynamically organize the blocks into an
approximately one-dimensional chain.  Their magnitudes should then
determine the proper separation between neighboring fuzzy spheres.
For the spherically symmetric saddle considered here, the resulting
link configuration must in addition preserve the diagonal $SO(3)$
symmetry.

To convert the link condensates into an effective notion of radial
distance, let us consider a block-diagonal test field that is constant
on each fuzzy sphere,
\be
\Phi
=
\bigoplus_s \phi_s\,{\bf 1}_{N_s},
\ee
so that variations of $\Phi$ probe only the radial direction.
The diagonal blocks commute with $\Phi$, whereas the off-diagonal
blocks give
\be
 [ {\mathbb X}_a,\Phi]_{s,s+1}
 =
 (\phi_{s+1}-\phi_s)W_a^{(s)}.
 \ee
 The commutator with the matrix background provides the natural notion
of a derivative in the matrix geometry.  We therefore probe the
effective radial geometry through the quadratic form
\be \label{EX}
{\cal E}_X[\Phi]
=
\frac12\sum_a
\Tr\!\left(
[\mathbb X_a,\Phi]^\dagger
[\mathbb X_a,\Phi]
\right),
\ee
which is the matrix analogue of the continuum gradient energy.
As a result, we obtain
\be \label{matrix-radial-energy}
 {\cal E}_{X}[\Phi]
 =
 \sum_s {\cal K}_s
 |\phi_{s+1}-\phi_s|^2 ,
\ee
where
\be \label{Ks}
 {\cal K}_s
 :=
 \sum_a
 \operatorname{Tr}
 W_a^{(s)\dagger}W_a^{(s)} .
\ee
Thus the link condensate determines a radial stiffness between the
neighboring blocks.

For a scalar field on a spherically foliated continuum geometry
\be
 ds_{\rm spatial}^2
 =
 d\rho^2+r^2(\rho)d\Omega_2^2 ,
\ee
the corresponding kinetic term is given by
\be
 {\cal E}_{\rm geom}
 =
 \frac12
 \int d\rho\,4\pi r^2(\rho)
 \left(\frac{d\phi}{d\rho}\right)^2 .
\ee
On a radial lattice this has the form
\be \label{continuum-radial-energy}
 {\cal E}_{\rm geom}
 \simeq
 2\pi
 \sum_s
 \frac{r_{s+1/2}^2}{\Delta\rho_s}
 |\phi_{s+1}-\phi_s|^2 ,
\ee
where
\be
 \Delta\rho_s:=\rho_{s+1}-\rho_s,
 \qquad
 r_{s+1/2}:=\frac12(r_s+r_{s+1}).
 \ee
Matching \eq{matrix-radial-energy} to
\eq{continuum-radial-energy} therefore allows the proper radial
spacing to be extracted from the link condensates.  In general, there
may be a single overall normalization ${\cal Z}$ relating the matrix
and continuum quadratic forms.  Defining it by
\be
{\cal E}_{\rm geom}={\cal Z}\,{\cal E}_{X},
\ee
one obtains
\be \label{rho-from-links}
\Delta\rho_s
=
\frac{2\pi r_{s+1/2}^2}
{{\cal Z}\,{\cal K}_s}.
\ee
The normalization ${\cal Z}$, which may carry dimensions depending
on the matrix conventions, is common to all links and must
be fixed independently from the normalization of the matrix
kinetic operator, rather than by fitting the emergent geometry.
The radial positions are then reconstructed as
\be
 \rho_s
 =
 \rho_1+\sum_{u=1}^{s-1}\Delta\rho_u .
\ee
The constant $\rho_1$ is fixed by the choice of radial origin.
If the innermost $N_1=1$ block represents the regular center, we set
$\rho_1 =0$.

The important point is that the two ingredients of the geometry are
now determined independently:
\be \label{matrix-two-data}
 N_s
 \quad\longrightarrow\quad
 r_s,
 \qquad\qquad
 W_a^{(s)}
 \quad\longrightarrow\quad
 \Delta\rho_s .
\ee
In principle, the matrix configuration predicts
both $r_s$ and the proper radial separation between successive
spheres.
This leads to a direct consistency test of the de Sitter geometry.
From the matrix data we may define
\be \label{Fs-matrix}
 F_s
 :=
 \left(
 \frac{r_{s+1}-r_s}{\Delta\rho_s}
 \right)^2 .
\ee
In the continuum limit this reconstructs the metric function through
\be
 ds_{\rm spatial}^2
 \simeq
 \frac{dr^2}{F(r)}
 +
 r^2d\Omega_2^2 .
\ee
For de Sitter space the prediction is
\be \label{dS-F-test}
 F_s
 \longrightarrow
 1-\frac{r_{s+1/2}^2}{R^2}.
\ee
Equivalently, the dimensionless quantity
\be \label{Ds}
 {\cal D}_s
 :=
 \left(
 \frac{r_{s+1}-r_s}{\Delta\rho_s}
 \right)^2
 +
 \frac{r_{s+1/2}^2}{R^2}
\ee
should satisfy
\be
 {\cal D}_s\longrightarrow 1
\ee
in the large-$N$ limit.  This constitutes a
nontrivial test where the second term of $\cD_s$
is determined by the representation sizes alone, whereas the first
involves the link condensates as well.


Note that
there is no need to introduce an independent time coordinate for the
interior matrix system.  The entire matrix model evolves with the same
time $t$ used for the fuzzy-sphere horizon and the asymptotically
normalized Schwarzschild exterior.  The canonically normalized
static-patch de Sitter time is instead
\be
 T=Ct,
\ee
where $C$ is the soldering parameter determined by the junction
condition.  The construction above addresses the emergence of the
spatial radial geometry.  A complete microscopic derivation of the
four-dimensional de Sitter metric would additionally require showing
that the temporal part of the low-energy matrix dynamics reproduces
the corresponding de Sitter redshift factor.

At present, this construction should therefore be regarded as a matrix
realization, together with a concrete consistency test, of the geometry
selected by the junction condition rather than as an independent
derivation of de Sitter space from the matrix equations of motion.
Microscopically, the natural quantities to be determined by the matrix
dynamics are the representation sizes $N_s$ and the condensates
$W_a^{(s)}$ of the neighboring link fields.  The former determine the
angular radii, while the latter determine the emergent radial
separations.  The gravitational analysis above supplies the target
configuration that such a matrix saddle should reproduce.  This
provides a possible concrete route by which gravity and spacetime may
emerge from the matrix degrees of freedom.

There are familiar precedents for this type of emergence in string
theory.  A fundamental string ending on a D-brane can be represented
as an electric BIon spike of the Abelian Born--Infeld world-volume
theory, while a D1-brane ending on a D3-brane appears as a magnetic
monopole spike
\cite{Callan:1997kz,Gibbons:1997xz}.  In the non-Abelian description,
analogous configurations are realized as fuzzy funnels, in which
matrix-valued world-volume fields reconstruct the geometry of a
higher-dimensional brane
\cite{Diaconescu:1996rk,Constable:2001ag}, or in
M-theory \cite{Basu:2004ed,Chu:2009iv}. In these examples an
extended bulk object emerges from degrees of freedom intrinsic to a
lower-dimensional matrix or gauge theory.  The construction proposed
here is similar in spirit: the radial bulk geometry is encoded in the
representation structure and link condensates of the microscopic
matrix degrees of freedom.

\section{Discussion}

In this article, we find that the quantum tadpole of the microscopic
black-hole horizon model has a direct geometrical consequence. In
isolation, the tadpole generates a positive area-conjugate
microscopic response and therefore energetically favors contraction
of the fuzzy sphere. However, once the horizon is embedded
in the complete gravitating system, the relevant condition is
stationarity of the combined horizon--gravity system. We find that the
nonvanishing tadpole determines, through the junction condition, the
gravitational background required for a self-consistent stationary
configuration. A de Sitter interior is obtained. In this respect, the
mechanism is close in spirit to the Fischler--Susskind mechanism.

The appearance of a de Sitter interior invites a possible
cosmological interpretation.
Since its curvature radius is fixed by
the black-hole horizon radius, $L=R=N\ell_P$, its Hubble scale and
cosmological constant are
\be
H_{\rm dS}=\frac{1}{N\ell_P},
\qquad
\Lambda\ell_P^2=\frac{3}{N^2}.
\ee
Thus a macroscopic de Sitter region automatically exhibits a large
hierarchy between the curvature and Planck scales.
If one tentatively identifies $H_{\rm dS}$ with the observed
dark-energy scale
\be
H_\Lambda=H_0\sqrt{\Omega_\Lambda},
\ee
the observed hierarchy corresponds parametrically to
\be
N\sim10^{61},
\qquad
\Lambda\ell_P^2\sim10^{-122}.
\ee
In this sense, the smallness of the de Sitter curvature may be viewed
as a consequence of the large matrix size of the quantum black-hole horizon.
We remark that this observation does not by itself
address the origin of dark matter
or the detailed cosmological evolution, which would require additional
degrees of freedom beyond the vacuum de Sitter background.

An interesting feature of our junction is the appearance of
the large relative lapse factor $C$.  This factor is not an intrinsic
parameter of the de Sitter geometry.
Rather, it specifies the relative lapse of
the interior metric with respect to the asymptotically normalized
Schwarzschild time in the exterior.  The normalization of the common
Killing vector is kept fixed.  What the null-horizon junction allows is
that the finite limiting value of the ratio $f/h$ need not agree on
the two sides of the horizon.  The parameter $C$ is therefore an
additional metric datum of the junction, and in the present model its
value is fixed by the microscopic horizon dynamics as
\be
 C\sim N
\ee
at large $N$, up to an $N$-independent numerical coefficient.
This is an intriguing relation. 
If $T=Ct$ denotes the canonically normalized static time of the de
Sitter interior, a frequency $\omega_{\rm BH}$ defined with respect to the
asymptotically normalized Schwarzschild time on the black hole side
corresponds to
\be
\omega_{\rm dS}=\frac{\omega_{\rm BH}}{C}.
\ee
In particular, a frequency of Planckian order with respect to
the Schwarzschild time is mapped to the de Sitter curvature scale:
\be \label{filter}
\omega_{\rm BH} \sim\ell_P^{-1}
\qquad\Longleftrightarrow\qquad
\omega_{\rm dS}\sim H_{\rm dS}.
\ee
Whether these relations admit a genuine cosmological interpretation,
and whether a slowly evolving quantum horizon could provide boundary
conditions for a quasi-de Sitter, inflationary-like cosmology, are
interesting questions for future study.

We remark that the emergence of a de Sitter interior is 
conceptually different from phenomenological de Sitter-core
and non-singular black-hole constructions
\cite{Frolov:1989pf,Dymnikova:1992ux,Hayward:2005gi}, as well as
gravastar-type models \cite{Visser:2003ge,
  Mazur:2004fk,Mazur:2015kia,Beltracchi:2018ait}.
In such constructions, the stress-energy required to support the
regular core or the junction is typically introduced
phenomenologically; in gravastar models this generally includes a
shell with an assumed stress tensor or equation of
state. Here the contractive
microscopic  response is instead supplied by
the same microscopic matrix degrees of freedom that define
the quantum horizon.
No new shell matter is introduced for the purpose of
supporting the de Sitter geometry.  In this sense, the non-singular
interior arises as the gravitational backreaction required by the
quantum dynamics of the horizon itself.

Previous applications of our quantum horizon model have utilized the
fermionic sector in essential ways in microscopic state counting
\cite{Chu:2024qil,Chu:2024edh}, Hawking tunneling \cite{Chu:2026wyh},
and the dynamical membrane
paradigm \cite{Chu:2026vzj,Chu:2026qom}.  The results of the present paper
establish a first
important role for the bosonic fluctuations, complementary to that of
the fermionic degrees of freedom:
the particular bosonic scalar (1,0) fluctuation mode controls
a simple component of the bulk gravitational response and helps to fix
the geometry in the interior of the horizon.  It would be interesting
to understand this result directly from the matrix-model point of view,
for example through a three-block analysis of the matrix quantum mechanics.

It is also natural to ask whether the remaining bosonic modes encode
more general gravitational responses.  An interesting extension is to
non-spherical horizon fluctuations.  Such modes are expected to couple
to bulk gravitational perturbations with the corresponding angular
momentum and parity and may therefore act as horizon sources for the
Regge--Wheeler and Zerilli sectors \cite{Regge:1957td,Zerilli:1970se}
and consequently
excite the black-hole quasinormal modes.
Establishing the precise map between the matrix
fluctuation spectrum and the exterior quasinormal response would
provide a concrete way to investigate whether the microscopic
structure of the quantum horizon can leave an imprint on
gravitational-wave observables.

Finally, we remark that the de Sitter interior is selected here by
imposing a static, spherically symmetric and regular interior vacuum
solution of the Einstein equations. These conditions are naturally
expected to describe the simplest gravitational configurations
compatible with the quantum mechanics. It would be interesting to understand
whether a larger class of quantum-corrected interiors is possible, and
how such configurations would emerge from the microscopic dynamics.


\section*{Acknowledgments}
  
 We thank  Ian Low for discussions and comments.
We acknowledge the support of this work by NCTS, the National Science and
Technology Council of Taiwan for the grant 113-2112-M-007-039-MY3, and
the National Tsing Hua University 2025 Talent Development Fund for a
TSAI WANG, YUAN-YANG Distinguished Talent Chair Professorship.


\bibliographystyle{utphys}
\bibliography{references}  
\end{document}